\documentclass{article}
\usepackage{amsmath}
\usepackage{amssymb}
\usepackage{graphicx}
\usepackage{verbatim}
\usepackage{ifthen}
\usepackage{epsfig}
\usepackage{setspace}
\usepackage{a4wide}
\newcounter{fig}   \newcommand{\lbfig}[1]{\refstepcounter{fig}
\label{#1} }

\newcommand{\Tr}{{\rm Tr}}

\newcommand{\bea}{\begin{eqnarray}}
\newcommand{\eea}{\end{eqnarray}}
\newcommand{\be}{\begin{equation}}
\newcommand{\ee}{\end{equation}}

\newcommand{\re}[1]{(\ref{#1})}

\begin{document}
\title{
Gravitating sphalerons in the Skyrme model}
\author{
{\large Ya. Shnir} \\
{\small BLTP, JINR, Dubna, Russia}\\
{\small Institute of Physics, Carl von Ossietzky University Oldenburg, Germany}
}
\maketitle
\begin{abstract}
We construct self-gravitating axially symmetric sphaleron solutions of the 3+1 dimensional Skyrme model
coupled to Einstein gravity. The solutions are static and
asymptotically flat, they are characterized by two integers $n$ and $m$, where $\pm n$ are the
winding numbers of the constituents and the second integer
$m$ defines type of the solution. These configuration correspond to the chains of
charge $n$ Skyrmions and charge $-n$ anti-Skyrmions placed along the axis of symmetry in alternating order.
We investigate the dependency of the masses of
the gravitating sphalerons on the gravitational coupling.
We find new chains of self-gravitating $|n| = 1$ Skyrmions-–anti-Skyrmions which
emerge at some critical non-zero value of the gravitational coupling and do not have flat space limit.
In contrast, the branches of self-gravitating $|n| \ge 2$ Skyrmion-–anti-Skyrmion chains
emerge from the corresponding flat space configurations. In both cases these
branches merge at some maximal
value of the effective gravitational coupling the branches of different type.
The branch of gravitating Skyrmion-–anti-Skyrmion pair extends all the way back to the limit
of vanishing coupling constant where solutions approach the corresponding generalised
Bartnik-McKinnon solutions. The upper branch of gravitating Skyrmion-–anti-Skyrmion--Skyrmion
chain exist up to some critical value of the gravitational coupling at which the chain becomes broken.
We further find that for
small values of the coupling constant on the upper branches, the solutions correspond to composite
systems, consisting of a scaled inner Einstein-Yang-Mills solution and outer Skyrmions which are separating from the
inner configuration.

\end{abstract}


\newpage

\section{Introduction}

There are many nonlinear classical field theories in flat spacetime which admit topologically stable soliton solutions.
These are particle-like, globally regular localised field configurations with finite energy.
Interesting examples in $d=3+1$ dimensions are monopoles in the
Yang-Mills-Higgs model \cite{'tHooft:1974qc,Polyakov:1974ek}, the solitons of the
Skyrme model \cite{Skyrme:1961vq} and knotted configurations of the Faddeev-Skyrme model
\cite{Faddeev}. The Skyrme model is nonlinear scalar $O(4)$ sigma-model, under certain assumptions it
can be derived from the expansion of the QCD low energy effective
Lagrangian in the large $N_c$ limit \cite{Witten}, then the topological charge of the
multisoliton configuration is set into
correspondence to the physical baryon number. Recently, some modifications of Skyrme
model was proposed to approach the topological bound  \cite{Adam:2010ds}
preserving the topological properties of the corresponding solitons.

Interestingly, in the flat space
both the Yang-Mills-Higgs theory and the Skyrme model admit classical sphaleron solutions
which represent monopole-anti-monopole pair in a static equilibrium \cite{Kleihaus:1999sx}, and
Skyrmion-anti-Skyrmion solution \cite{Krusch:2004uf}, respectively. Further,
there are axially symmetric generalizations of these solutions which represent chains of interpolating
Skyrmion--anti-Skyrmions \cite{Shnir:2009ct} and monopole--antimonopole chains \cite{KKS}, each carrying topological
charge $n$ in alternating order.
For an even number of constituents, these chains reside in the topologically trivial
sectors of both models. When the number of constituents is odd, the configurations represent deformations
of the axially symmetric multisoliton of degree $n$.

When gravity is coupled to the matter field, this has significant
effect on these solutions. It turns out there are hairy
black hole solutions of the Einstein-Skyrme theory \cite{Luckock:1986tr,Droz:1991cx,Bizon:1992gb}.
Historically, that was the first example of constructions of hairy black holes.
These solutions are stable, asymptotically flat and possess a regular horizon, furthermore, they may
be viewed as bound states of Skyrmions and Schwarzschild black holes \cite{Ashtekar}.
Axially-symmetric static solutions of the Einstein-Skyrme model with topological charge two were studied in
\cite{Sawado:2004yq}.
Recently, in \cite{Adam:2014dqa},
self-gravitating BPS Skyrmions were applied in description of bulk properties of neutron stars.

The globally regular gravitating Skyrmions in asymptotically flat space
were studied in \cite{Droz:1991cx,Bizon:1992gb}. It was shown that there are two
branches of solutions, one of which emerges smoothly from
the flat space  Skyrmion configuration. When the effective gravitational coupling constant
$\alpha^2=4\pi G F_\pi$, where $G$ is the Newton constant and $F_\pi$ is the pion decay constant,
is
increased from zero, this branch terminates at some critical value of the coupling,
beyond which gravity becomes too strong for self-gravitating Skyrmions to
persist. There it merges with a second branch, which extends all the
way back to vanishing coupling constant. Along this branch the mass of the gravitating Skyrmion rapidly increases and
the solution becomes unstable. Surprisingly, it was shown \cite{Bizon:1992gb}
that in the limit of vanishing coupling,
the gravitating Skyrmion approaches the lowest Bartnik-McKinnon (BM) solution of the
$SU(2)$ Einstein-Yang-Mills theory \cite{Bartnik:1988am}. This pattern is rather similar to
the branch structure of the gravitating monopole-antimonopole pair \cite{Kleihaus:2000hx}. Indeed, the effective
coupling constant of the Einstein-Yang-Mills-Higgs theory is proportional to the Higgs
vacuum expectation value $\eta$ and the square root of the
gravitational constant $G$. The evolution of the sphaleron solution along the second branch
may be considered as being obtained by decreasing the vacuum expectation value of the scalar field, in the
limit of $\eta \to 0$ the configuration smoothly approaches the same lowest mass spherically symmetric
BM solution.

The pattern of evolution of the self-gravitating
monopole-anti-monopole chains and vortex rings in Einstein-Yang-Mills-Higgs theory
is qualitatively similar to that of the monopole-antimonopole pair, the difference is that
on the upper branches, these solutions correspond to composite systems, consisting of a scaled
inner Einstein-Yang-Mills solution and an outer flat space Yang-Mills-Higgs solution \cite{Kleihaus:2004fh}.

The properties of gravitating Skyrmions were considered in many works, for example configurations with
discrete symmetry were investigated in \cite{Ioannidou:2006mg}, spinning gravitating Skyrmions were studied in
\cite{Ioannidou:2006nn}. Also modifications of the Einstein-Skyrme model with cosmological constant were investigated
in \cite{Shiiki:2005aq,Brihaye:2005an}.
In this paper we present globally regular gravitating axially-symmetric Skyrmion-anti-Skyrmion configurations.

Since the consistent consideration of the solitons with higher
number of constituents is related with complicated task of  numerical simulations,
we restrict our consideration to the case of the Skyrmion-anti-Skyrmion (S-A)
pair and Skyrmion-anti-Skyrmion-Skyrmion (S-A-S) chain with constituents of degrees
$n = 1,2$. We show that the general pattern of evolution of the configuration is similar to
the branch structure of the monopole-antimonopole chains in the Einstein-Yang-Mills-Higgs system which
links the corresponding flat space configurations and the BM solution.

\section{The model}

The Einstein--Skyrme model in asymptotically flat 3+1 dimensional space is defined by the
action
\begin{equation} \label{action}
S =  \int \left\{ \frac{R}{16\pi G}
+ L_{m}\right\} \sqrt{- g} d^4 x
\end{equation}
where the gravity part of the action is the usual Einstein--Hilbert
action with curvature scalar $R$, $g$ denotes the determinant
of the metric, $G$ is the gravitational constant and the matter part of the action is
given by the Skyrme Lagrangian
\be
\label{Skyrme}
L_{m} = \frac{F_\pi^2}{16}g^{\mu\nu}{\Tr}\left( L_\mu L_\nu\right) + \frac{1}{32 e^2}g^{\mu\nu}
g^{\rho\sigma} {\Tr} \left([L_\mu,L_\rho][L_\nu,L_\sigma]\right) +\frac{\mu_\pi^2 F_\pi^2}{8}\Tr
\left(U-1\right)
\ee
where $L_\mu = U^\dagger \partial_\mu U$ is the $su(2)$-valued left current, associated with
the $SU(2)$-valued Skyrme field $U = \sigma \cdot {\mathbb I} + i \pi^a \cdot \tau^a$.
Here the quartet of the fields $(\sigma, \pi^a)$ is restricted to the surface
of the unit sphere, $\sigma^2+ \pi^a \cdot \pi^a =1$ and
$F_\pi, e$ and $\mu_\pi$ are parameters of the Skyrme model. Explicitly, $F_\pi$
is the so-called pion decay constant, $e$ is dimensionless constant and $\mu_\pi$ is the
tree-level pion mass. Note that in the rescaled radial coordinate $r \to e r F_\pi/2 $ the action of the
Einstein-Skyrme theory becomes
\begin{equation}
\label{action2}
S =  \int \left\{ \frac{R}{\alpha^2}
+ \frac{1}{2} g^{\mu\nu}{\Tr}\left( L_\mu L_\nu\right)
+ \frac{1}{16} g^{\mu\nu}
g^{\rho\sigma} {\Tr} \left([L_\mu,L_\rho][L_\nu,L_\sigma]\right) + \mu^2 \Tr (U-1)
\right\} \sqrt{- g} d^4 x
\end{equation}
where $\mu=2\mu_\pi/(F_\pi e)$ is the rescaled mass parameter. Thus,
the only physical parameter of the Einstein-Skyrme theory
\re{action2} is the effective gravitational coupling constant
$\alpha^2=4\pi G F_\pi$ \cite{Bizon:1992gb}.

Variation of the action \re{action2} with respect to the metric $g_{\mu\nu}$ yields the rescaled Einstein equations
\be
R_{\mu\nu} -\frac12 R g_{\mu\nu} = \alpha^2 T_{\mu\nu}
\ee
where Skyrme stress-energy tensor is
\be
\label{energy}
\begin{split}
T_{\mu\nu} =& \Tr \left(\frac12 g_{\mu\nu} L_\alpha L^\alpha - L_\mu L_\nu \right)
+ \Tr \biggl(g_{\mu\nu}[L_\alpha,L_{\beta}][L^\alpha,L^{\beta}]\\
&- 4 g_{\alpha\beta} [L_\mu,L^{\alpha}]
[L_\nu,L^{\beta}] \biggr) + \mu^2 g_{\mu\nu} \Tr (U-1) \, .
\end{split}
\ee

In order to construct finite energy topologically non-trivial solutions of the model \re{action2}
the Skyrme field must approach the unit matrix at spacial boundary, $U\to 1$ as $r\to \infty$, thus the
field is a map $U : S^{3} \to S^{3}$,
which belongs to an equivalence class characterized by the homotopy group
$\pi_3(SU(2))=\mathbb{Z}$. The corresponding topological current is
\be
B_\mu = \frac{1}{24 \pi^2 \sqrt{-g}}\varepsilon_{\mu\nu\rho\sigma} \Tr (L^\nu L^\rho L^\sigma) \, .
\ee

To obtain gravitating static axially symmetric solutions,
we employ the usual Lewis-Papapetrou metric in isotropic coordinates:
\be
\label{metric}
ds^2=
  - f dt^2 +  \frac{m}{f} d r^2 + \frac{m r^2}{f} d \theta^2
           +  \frac{l r^2 \sin^2 \theta}{f} d\varphi^2\, ,
\ee
where the metric functions
$f$, $m$ and $l$ are functions of
the radial variable $r$ and polar angle $\theta$, only.
The $z$-axis ($\theta=0, \pi$) represents the symmetry axis.

Since we are only concerned with axially symmetric fields, the
$SU(2)$ valued Skyrme field $U({\bf r}) = \sigma \cdot {\mathbb I} + i \pi^a \cdot \tau^a$
can be parameterized as \cite{Krusch:2004uf,Shnir:2009ct}
\be
\label{ansatz}
\pi^1= \phi^1 \cos(n \varphi);\quad \pi^2=\phi^1 \sin(n \varphi);\quad \pi^3=\phi^2;\quad \sigma = \phi^3
\ee
where the triplet of scalar fields $\phi^a$ on unit sphere is a function only of radial variable $r$ and polar angle $\theta$.
 One can check
that the parametrization \re{ansatz} is consistent, i.e. the complete set of the field equations, which
follows from the variation of the action of the Einstein-Skyrme model, is compatible with
the equations on three functions $\phi^a$, which follow from variation of the reduced action on
the ansatz \re{ansatz}.

An equivalent parametrization of the axially-symmetric Skyrme field is given by
\be
\label{trig}
\phi^1 = \sin f \sin g;\qquad \phi^2 =  \sin f \cos g;\qquad \phi^3 = \cos f
\ee
where the functions $f(r,\theta)$ and $g(r,\theta)$ depend on the radial coordinate $r$ and polar angle $\theta$.
Imposing the boundary condition on the asymptotic value of the function $g(r,\theta)$ \cite{Shnir:2009ct}
\be
\lim_{r\to\infty} g(r,\theta) = m \theta
\ee
where $m$ is an integer, we can construct various multi-Skyrmion or
the Skyrmion–-anti-Skyrmion chain solution of given degree. Then the value of
$m$ corresponds to the number of the constituents of the configuration,
each of them can be identified with an individual charge $\pm n$ Skyrmion.
Evaluation of the corresponding baryon number yields
$B= \frac{n}{2} \left[1-(-1)^m\right]$, i.e. the case $m=1$ corresponds to the multi-Skyrmions,
in particular $n=1,m=1$ configuration is the usual spherically symmetric Skyrmion parameterized via
the function of radial variable $f(r)$.

The $m=2$ configuration is
the Skyrmion-anti-Skyrmion (S-A) pair \cite{Krusch:2004uf}, which is a deformation of the topologically trivial sector.
Note that in the absence of the potential term in the flat space only topological charge
pairs of $n \ge  2$ are found. As we will see, the gravitational interaction between the constituents
makes it possible for $n=1$ self-gravitating Skyrmion-–anti-Skyrmion pairs to exist.

The configurations with $m=3$ correspond to the
Skyrmion-anti-Skyrmion-Skyrmion chain (S-A-S)
\cite{Shnir:2009ct}. Thereafter we restrict our consideration to
the axially symmetric solutions with values $n=1,2$ and $m=1,2,3$.
Note
that the binding energy of the constituents of the Skyrmion-anti-Skyrmion chains is quite weak,
so the $|n| = 2$ chains in the flat space are very unstable with respect to perturbations \cite{Shnir:2009ct}.

Clearly, these configurations and monopole–-anti-monopole chain solutions of the Yang-Mills-Higgs theory
\cite{Kleihaus:1999sx,KKS} have many features in common.
In both cases there is an effective
interaction between the constituents which allows the sphaleron solution to exist, although
the nature of the interaction is different. In the Yang-Mills-Higgs theory it is an effective electromagnetic
interaction between the monopoles \cite{Shnir:2005te} while in the case of the Skyrme theory there is a dipole-–dipole
interaction between the Skyrmions \cite{Krusch:2004uf,Shnir:2009ct}. Since the latter interaction is much weaker than
the effective electromagnetic forces,
in the absence of the mass term the Skyrmion-anti-Skyrmion chains may exist in the flat space
only when each of the constituents carries charge $|n| \ge 2$. Clearly, coupling to gravity provides additional
attraction between the lumps which results in the existence of self-gravitating Skyrmion-anti-Skyrmion chains with
$|n|=1$. Thereafter we set the mass constant $\mu=0$.

The complete set of the field equations, which can be obtained from
variation of the action of the Einstein-Skyrme model \re{action2},
can be solved when we impose the boundary conditions on the fields
$\phi^a$ and make use of the parametrization of the metric
\re{metric}. Then the field equations reduce to a set of six
coupled partial differential equations, to be solved numerically.

As usually, the boundary conditions follow from the regularity on the symmetry axis
and symmetry requirements as well as
the condition of finiteness of the energy. In particular we have to take into account that
the asymptotic value of the Skyrme field is restricted to the vacuum and
the metric functions must approach unity at the spacial boundary.
Explicitly, we impose
\be
\begin{split}
\phi^{1}\biggl.\biggr|_{r \rightarrow \infty }\!\!\! & \rightarrow
0\,,~~~\phi^{2}\biggl.\biggr|_{r \rightarrow \infty
}\!\!\!\rightarrow 0\,,~~~\phi^{3}\biggl.\biggr|_{r \rightarrow
\infty }\!\!\!\rightarrow 1\, ,\\
f\biggl.\biggr|_{r \rightarrow \infty
}\!\!\! & \rightarrow  1\, ,  ~~~m\biggl.\biggr|_{r \rightarrow \infty
}\!\!\!\rightarrow 1\, ,
~~~l\biggl.\biggr|_{r \rightarrow \infty
}\!\!\!\rightarrow 1
\label{infty1}
\end{split}
\ee%
at infinity. For odd values of $m$ at the origin we impose
\begin{equation}
\begin{split}
\phi^{1}\biggl.\biggr|_{r \rightarrow 0 }\!\!\! & \rightarrow
0\,,~~~\phi^{2}\biggl.\biggr|_{r \rightarrow 0
}\!\!\!\rightarrow 0\,,~~~\phi^{3}\biggl.\biggr|_{r \rightarrow
0 }\!\!\!\rightarrow -1\, ,\\
\partial_r f\biggl.\biggr|_{r \rightarrow 0
}\!\!\! & \rightarrow 0\,,  ~~~\partial_r m \biggl.\biggr|_{r \rightarrow 0
}\!\!\!\rightarrow 0\,,
~~~\partial_r l\biggl.\biggr|_{r \rightarrow 0
}\!\!\!  \rightarrow 0 \, .
\label{origin}
\end{split}
\end{equation}%
For the S-A pair the boundary conditions on the Skyrme field are different \cite{Krusch:2004uf,Shnir:2009ct}
\be
\phi^{1}\biggl.\biggr|_{r \rightarrow 0 }\!\!\!  \rightarrow
0\,,~~~\phi^{2}\biggl.\biggr|_{r \rightarrow 0
}\!\!\!\rightarrow 0\,,~~~\phi^{3}\biggl.\biggr|_{r \rightarrow
0 }\!\!\!\rightarrow 0\,
\ee
since the position of the components corresponds to the points in space where the
third component of the field $\phi^3$ is equal to its anti-vacuum value $-1$.
\begin{figure}[hbt]
\lbfig{f-1}
\begin{center}
\includegraphics[height=.32\textheight, angle =-90]{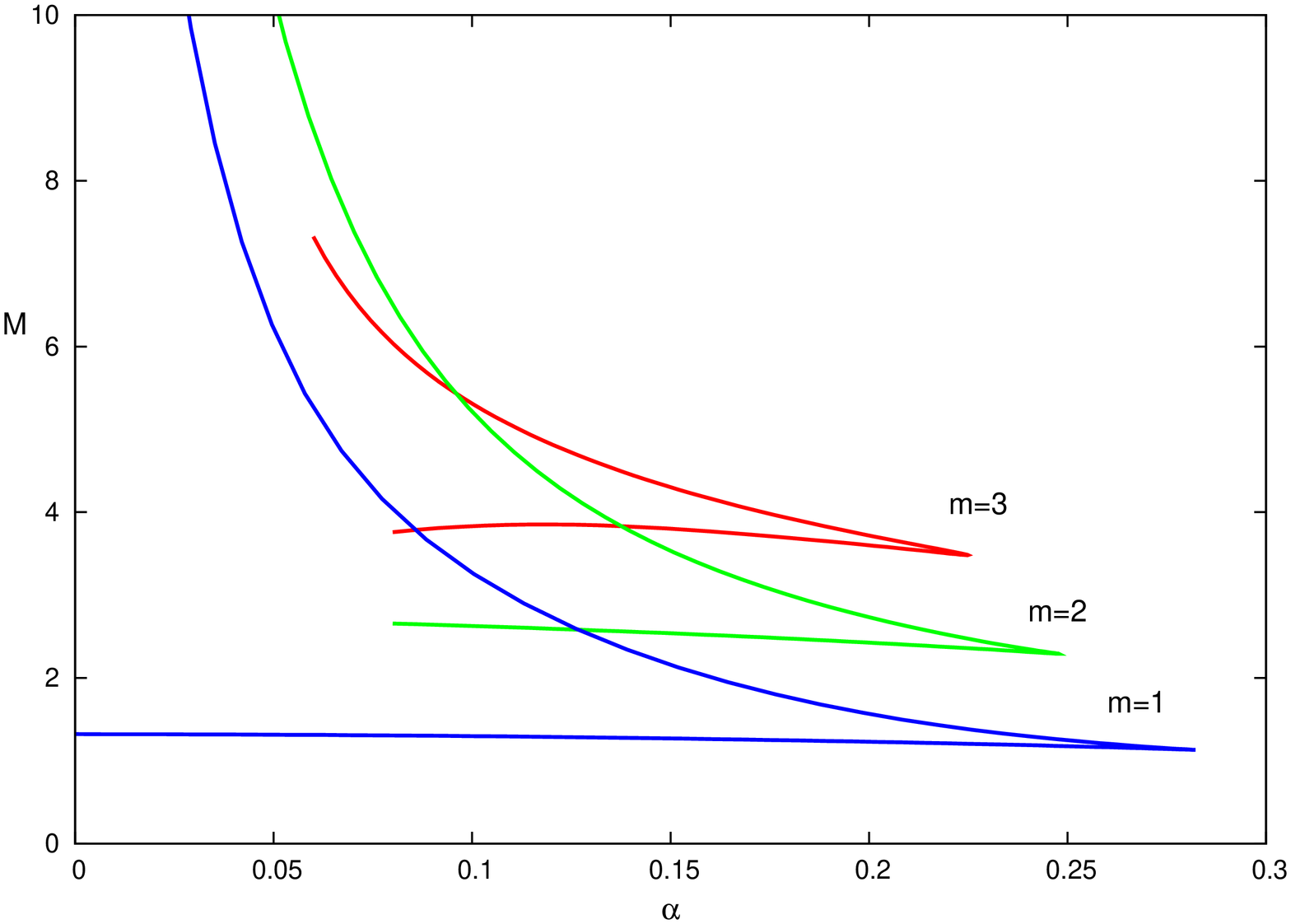}
\includegraphics[height=.32\textheight, angle =-90]{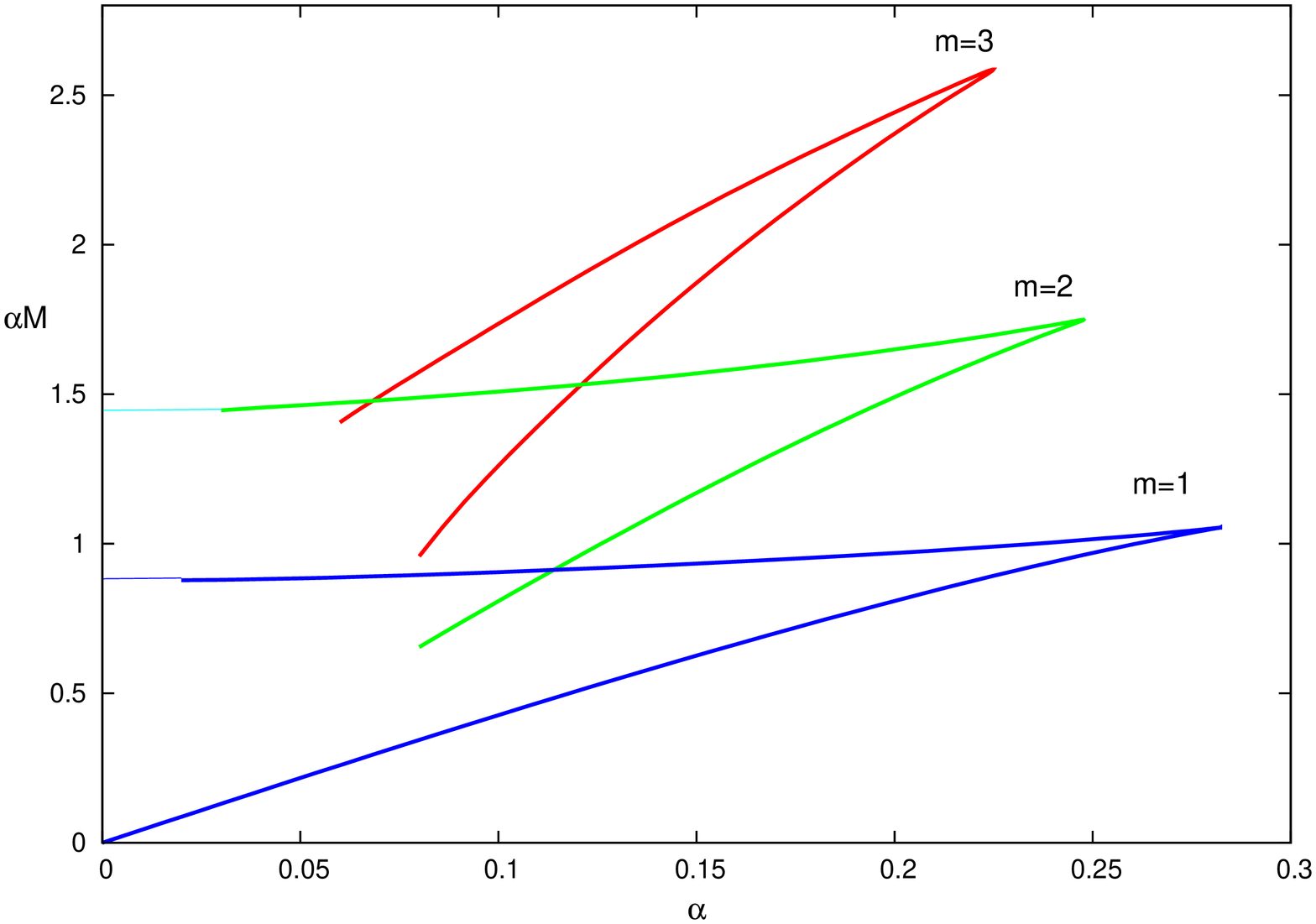}
\end{center}
\caption{\small
The mass $M$ (left), and the scaled mass $\hat M$ (right) of the charge 1 gravitating
Skyrmion and the Skyrmion–-anti-Skyrmion chains for $n = 1$ , $m = 1,2,3$
are shown as functions of the coupling constant $\alpha$. The thin lines extend the Skyrmion
curves of the scaled mass to the mass of the corresponding generalized Bartnik-McKinnon solution.
}
\end{figure}

The condition of regularity of the functions on the symmetry axis yields
\begin{equation}
\begin{split}
\phi^{1}\biggl.\biggr|_{\theta \rightarrow 0,\pi }\!\!\! & \rightarrow
0\,,~~~\partial_\theta \phi^{2}\biggl.\biggr|_{\theta \rightarrow 0,\pi }
\!\!\!\rightarrow 0\,,~~~\partial_\theta \phi^{3}\biggl.\biggr|_{\theta \rightarrow 0,\pi }
\!\!\!\rightarrow 0\, ,\\
\partial_\theta f\biggl.\biggr|_{\theta \rightarrow 0,\pi }
\!\!\! & \rightarrow 0\,,  ~~~\partial_\theta m \biggl.\biggr|_{\theta \rightarrow 0,\pi }
\!\!\!\rightarrow 0\,,
~~~\partial_\theta l\biggl.\biggr|_{\theta \rightarrow 0,\pi }\!\!\!  \rightarrow 0
\label{bc4a}
\end{split}
\end{equation}%
To satisfy the condition of regularity $m(r,0)=l(r,0)$ we introduce the auxiliary function
$g(r,\theta) = l(r,\theta)/m(r,\theta)$
with the boundary conditions $g(0,\theta)=g(\infty,\theta)=g(r,0)=1; \, \partial_\theta g(r,\pi/2 )= 0)$. We check this
condition as a test for correctness of our numerical results.

\section{Numerical results}
The numerical calculations are mainly performed on an equidistant grid
in spherical coordinates $r$ and $\theta$, employing the compact radial coordinate $x= r/(1+r) \in
[0:1]$ and $\theta \in [0,\pi]$.
To find solutions of the Euler-Lagrange equations which follow from the rescaled action \re{action2} and
depend parametrically on the effective gravity constant $\alpha$,
we used the software package CADSOL based on the Newton-Raphson algorithm \cite{schoen}.
This code solves a given system of nonlinear partial differential equations
subject to a set of boundary conditions on a rectangular domain.
Typical grids we used have sizes $90 \times 70$.
The resulting system is solved iteratively until convergence is achieved.

Apart from some initial guess for the  solution, CADSOL requires also
 the Jacobian matrices for the equations with respect to  the unknown
 functions and their first and second derivatives, and the boundary conditions.
This software package provides also error estimates for each function,
which allows to judge the quality of the computed solution.
The relative errors of the solutions we found, are of order of $10^{-3}$ except the S-A-S chains for those they are as
large as $10^{-2}$.
We also introduce an additional Lagrangian multiplier to constrain the field to the surface of unit
sphere.

\begin{figure}[hbt]
\lbfig{f-2}
\begin{center}
\includegraphics[height=.32\textheight, angle =-90]{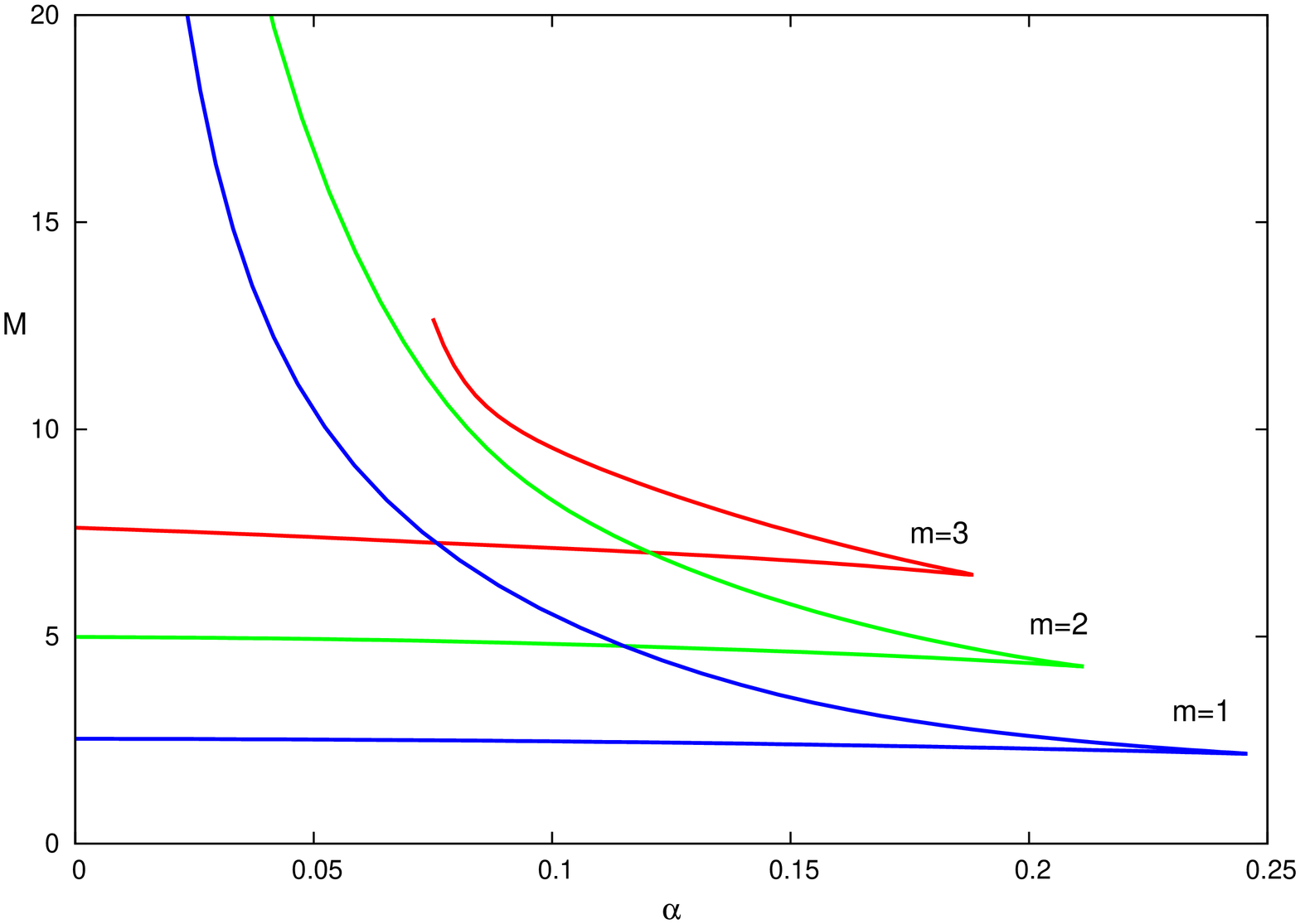}
\includegraphics[height=.32\textheight, angle =-90]{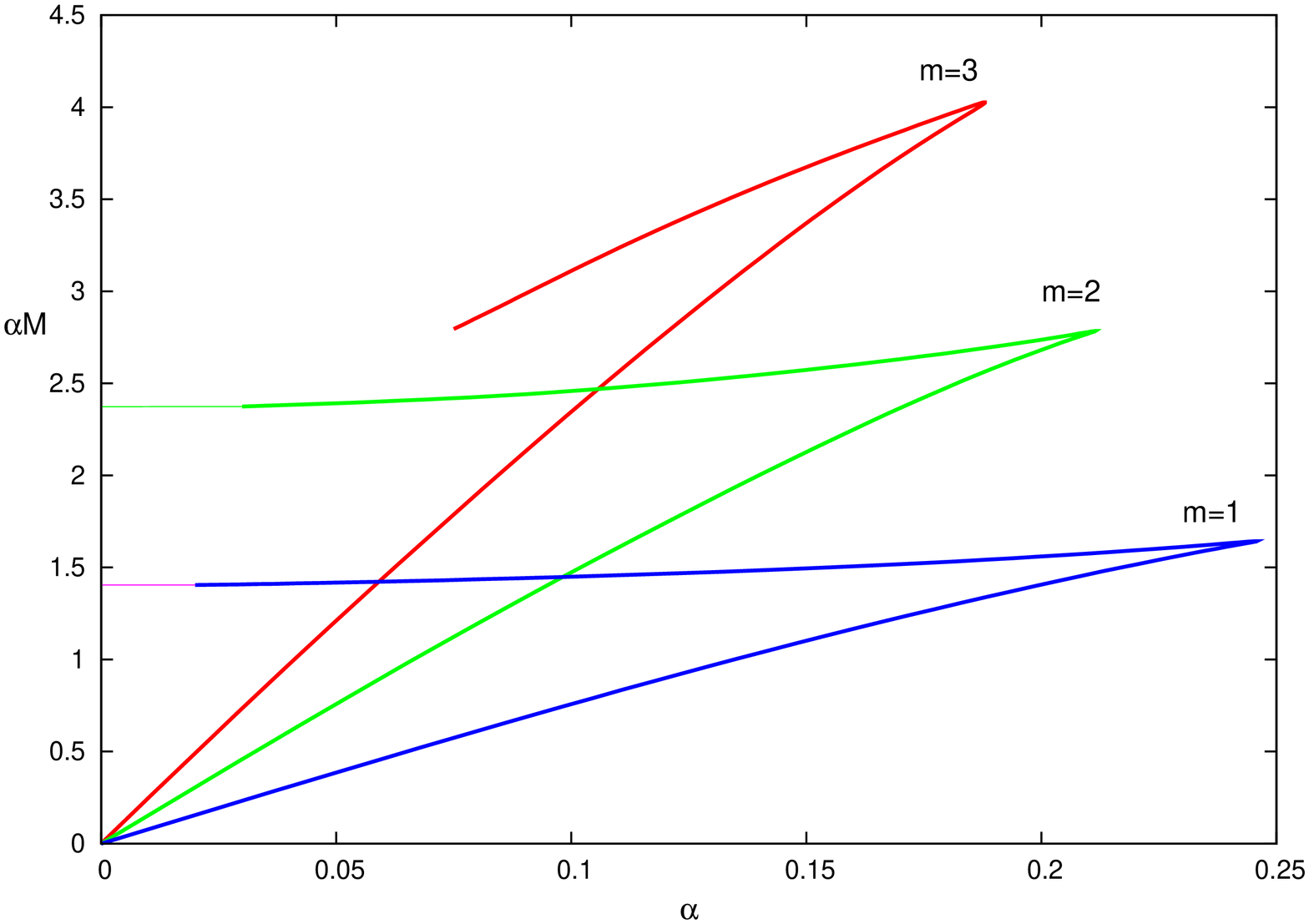}
\end{center}
\caption{\small
The mass $M$ (left), and the scaled mass $\hat M$ (right) of the charge 2 gravitating
Skyrmion and the Skyrmion–-anti-Skyrmion chains for $n = 2$ , $m = 1,2,3$
are shown as functions of the coupling constant $\alpha$. The thin lines extend the Skyrmion
curves of the scaled mass to the mass of the corresponding generalized Bartnik-McKinnon solution.
}
\end{figure}

\begin{figure}[hbt]
\lbfig{f-3}
\begin{center}
\includegraphics[height=.32\textheight, angle =-90]{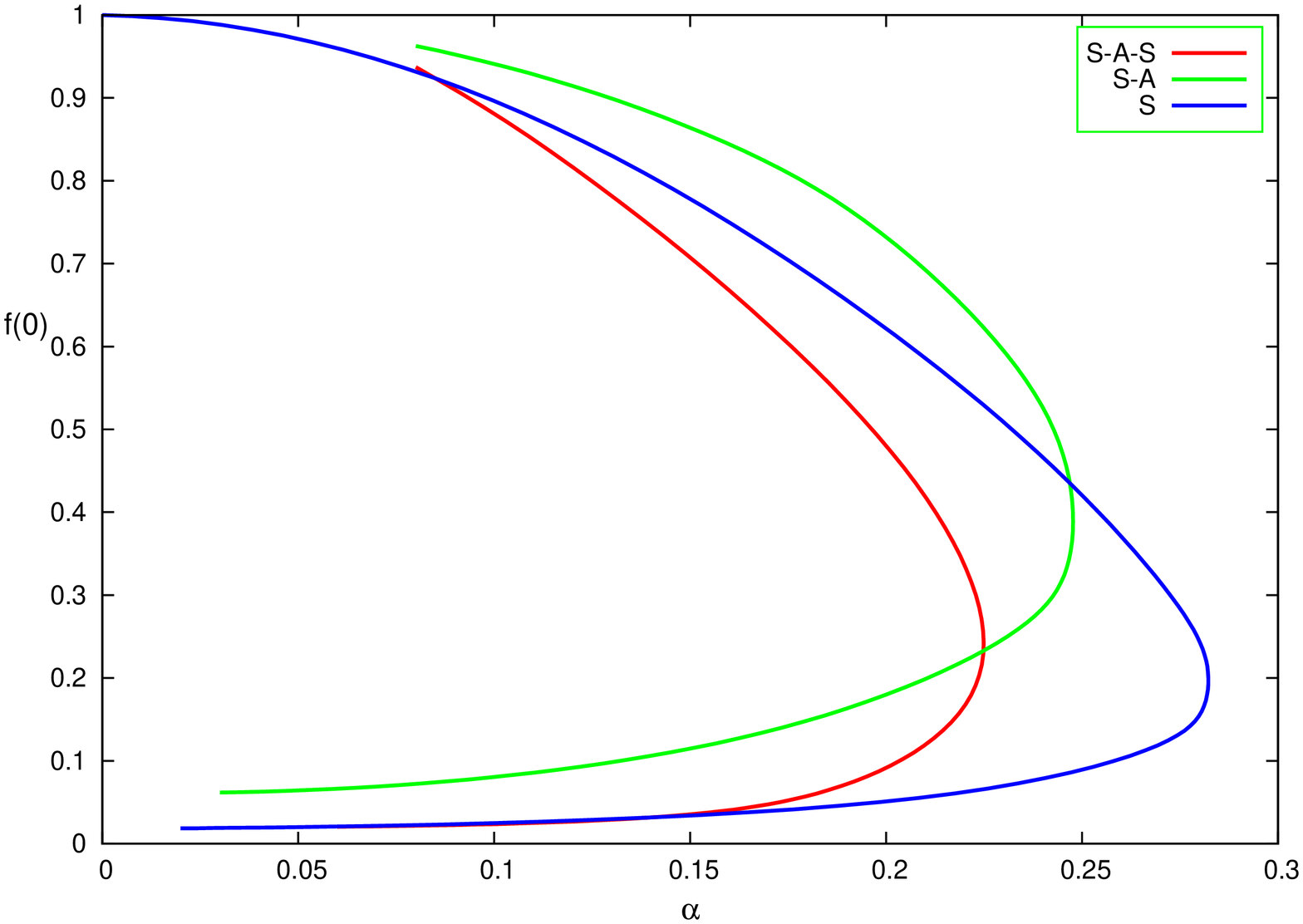}
\includegraphics[height=.32\textheight, angle =-90]{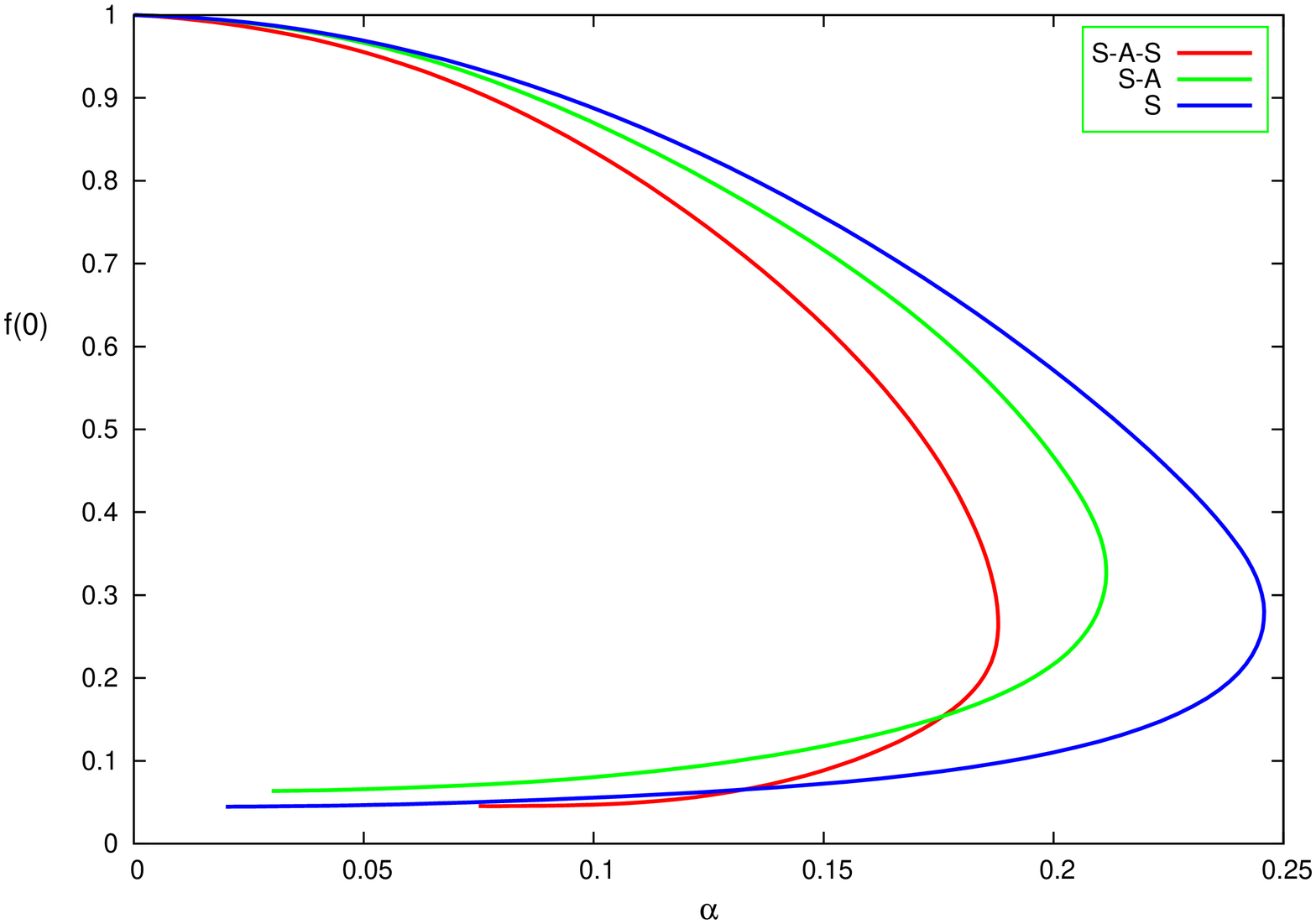}
\end{center}
\caption{\small
The values of the metric functions $f(0)$ at the origin
are shown as functions of the gravitational coupling constant $\alpha$ for
the gravitating Skyrmion and
the Skyrmion–-anti-Skyrmion chains with $n = 1$ (left) and $n=2$ (right) with $m = 1,2,3$.
}
\end{figure}

\begin{figure}[t]
\lbfig{f-4}
\begin{center}
\includegraphics[height=.32\textheight, angle =-90]{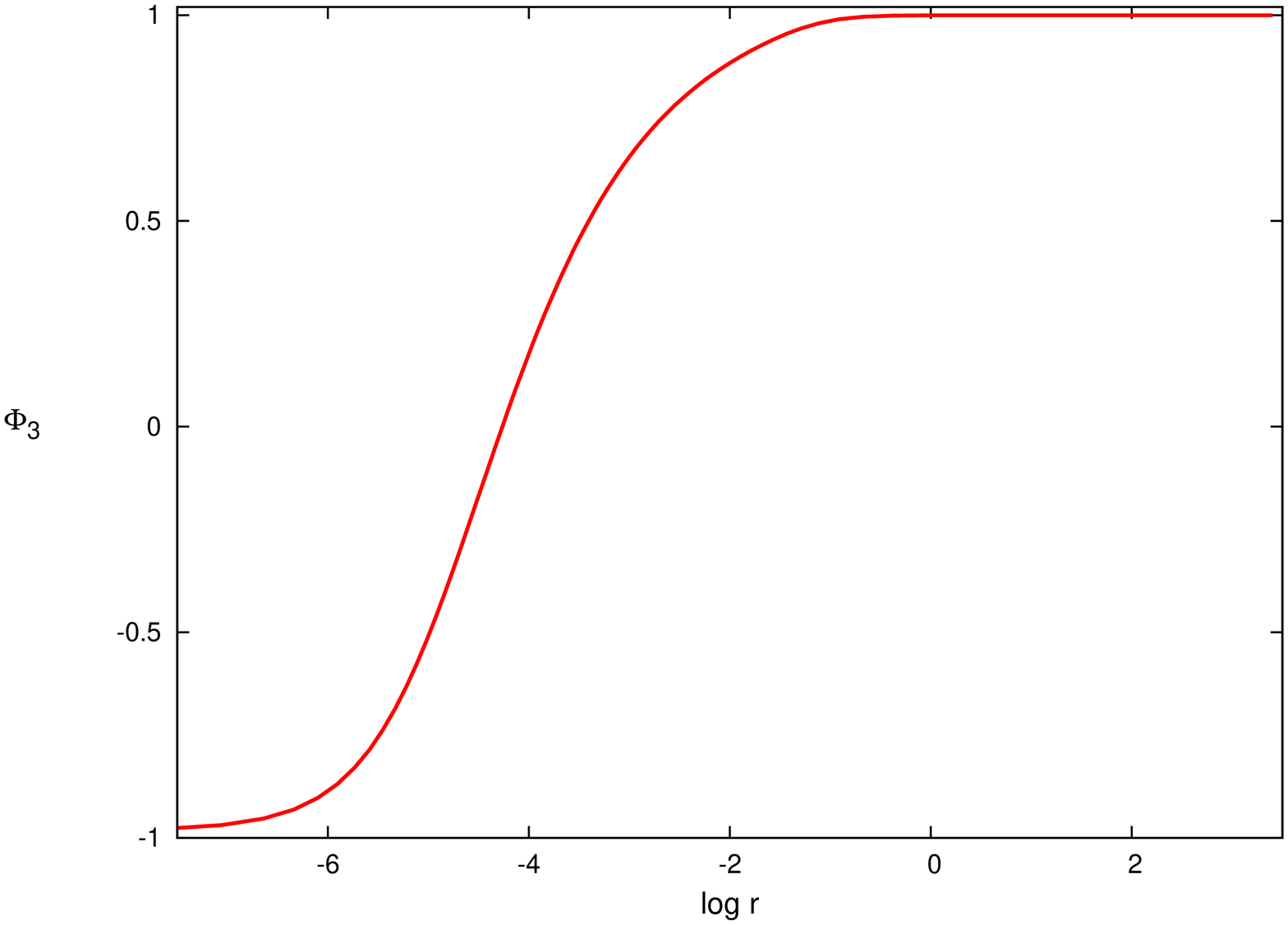}
\includegraphics[height=.32\textheight, angle =-90]{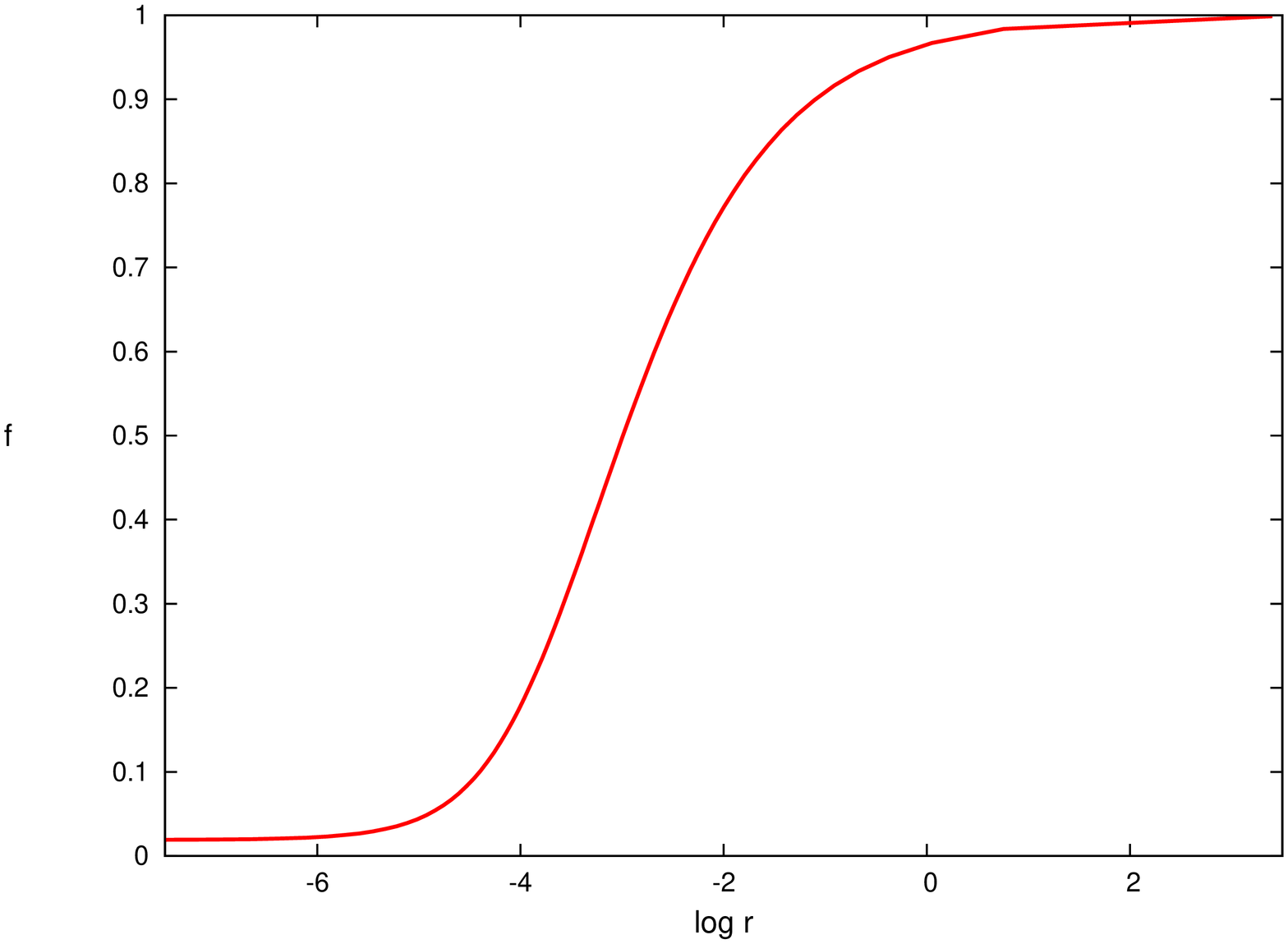}
\includegraphics[height=.32\textheight, angle =-90]{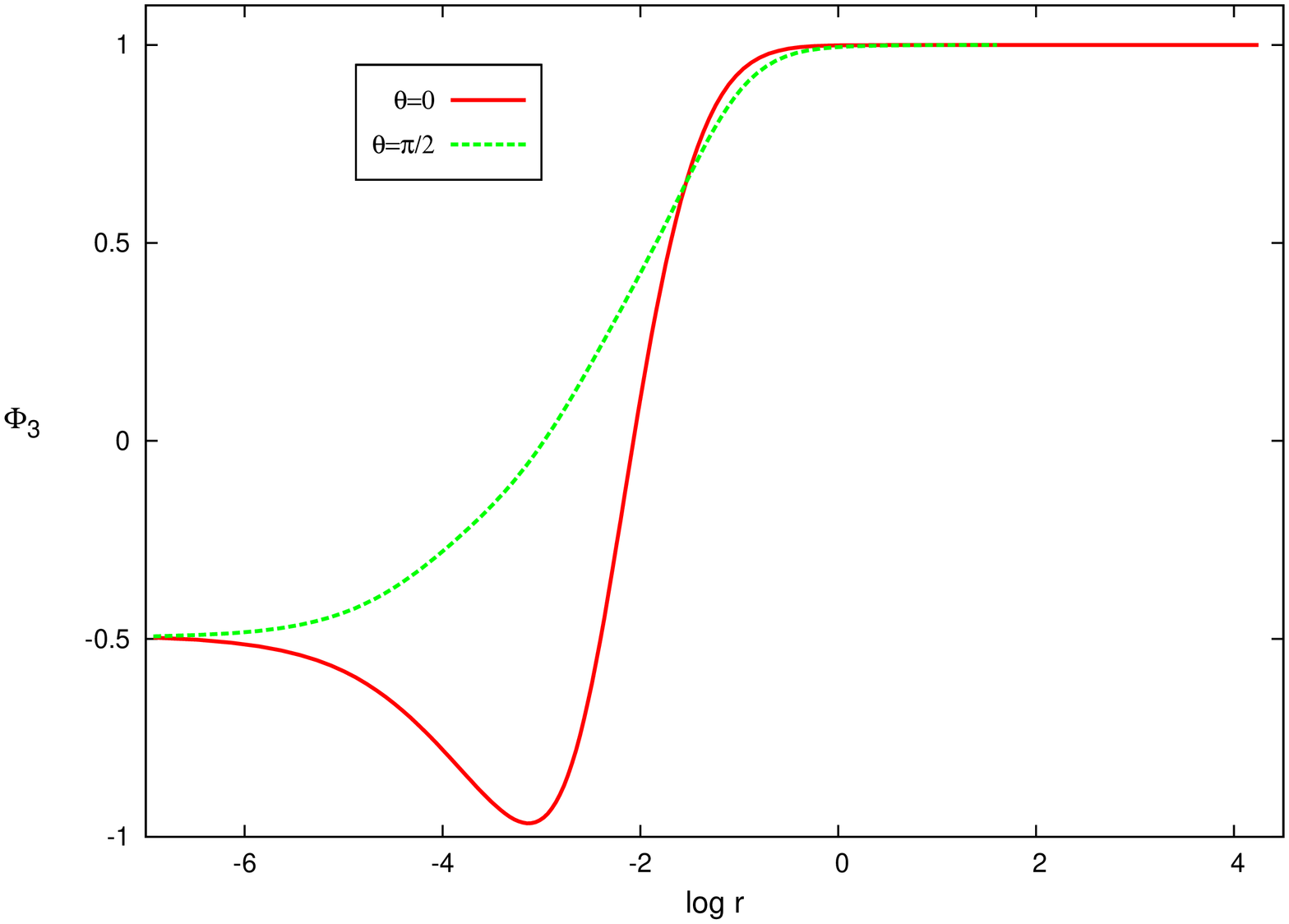}
\includegraphics[height=.32\textheight, angle =-90]{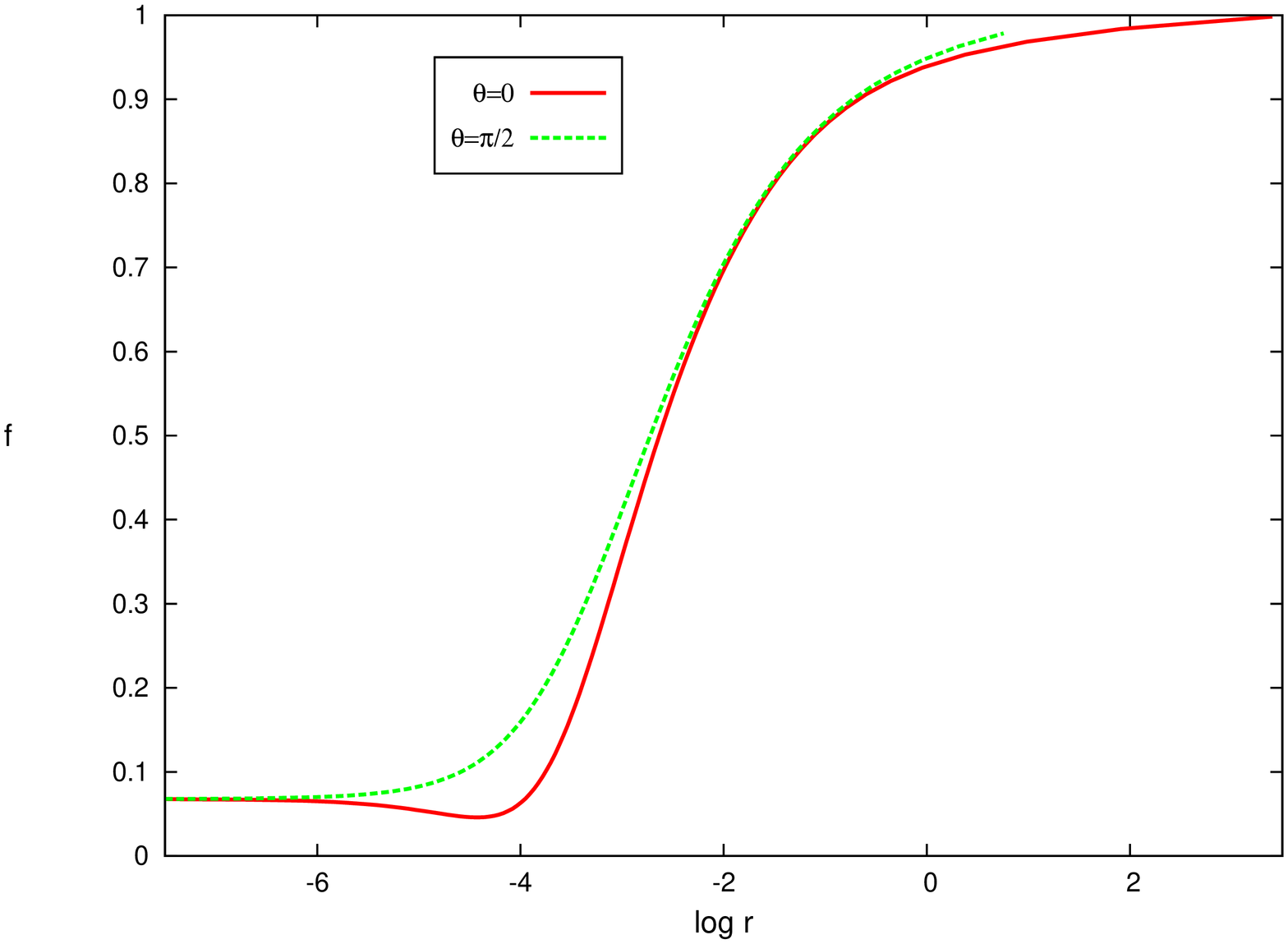}
\includegraphics[height=.32\textheight, angle =-90]{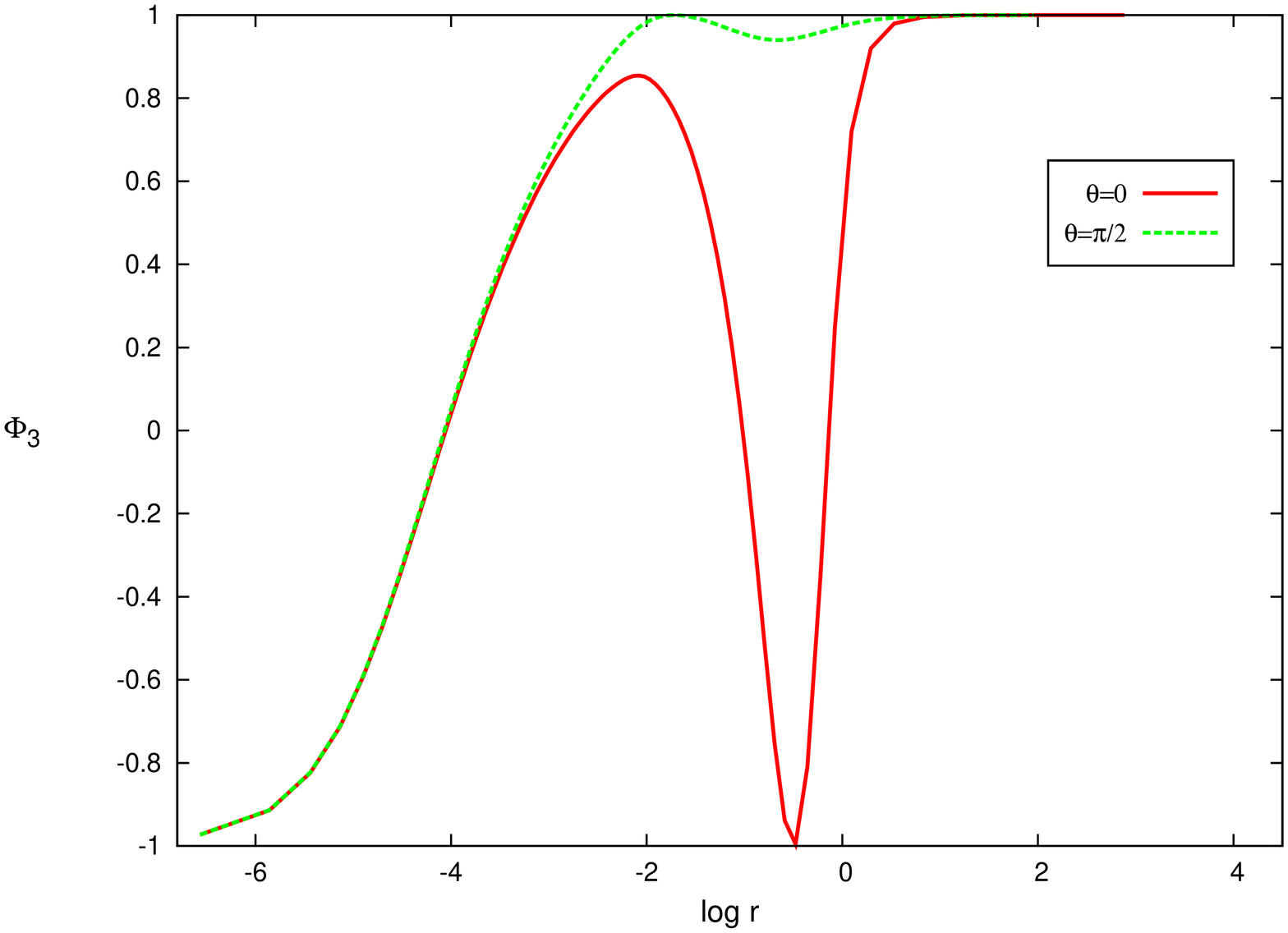}
\includegraphics[height=.32\textheight, angle =-90]{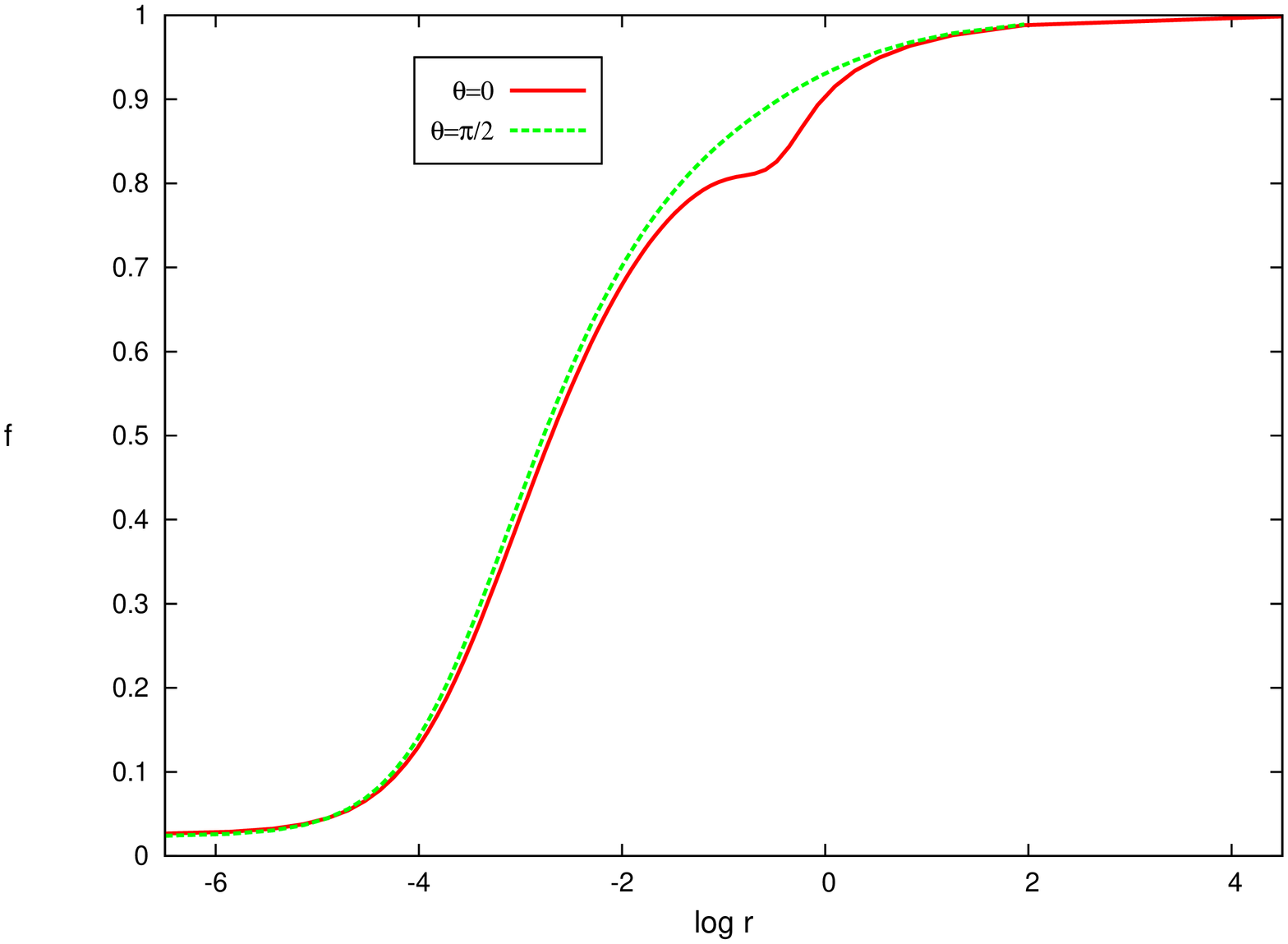}
\end{center}
\caption{\small
The  metric function $f(r,\theta)$ (right column) and the third components of the field of the
gravitating spherically symmetric Skyrmion and the Skyrmion–-anti-Skyrmion chains
with components of charge $n=1$ (left column) on the upper branch are shown in logarithmic radial coordinate
$\log r$ at $\alpha^2 = 0.009$ (for $m=1,2$) and at $\alpha^2 = 0.01$ (for $m=3$.
}
\end{figure}

Note that dimensionless gravitational coupling constant $\alpha^2=4\pi G F_\pi$ vanishes if
(i) the Newton constant $G \to 0$, or, (ii) $F_\pi \to 0$. In the
former case one may recover the sphaleron solitons of the Skyrme
model in the flat space, in the second case the
sigma-model term in the action \re{action} is vanishing. Thus, similar to
the case of the self-gravitating monopole-antimonopole systems in the asymptotically flat Einstein-Yang-Mills-Higgs theory
\cite{Kleihaus:2000hx,Kleihaus:2004fh} and Yang-Mills sphalerons in the AdS$_4$ spacetime \cite{Kichakova:2014fta},
there are two branches of solutions of the Einstein-Skyrme model \cite{Bizon:1992gb,Volkov:1989fi,Ioannidou:2006nn}.

The dependence of the spherically symmetric $n=1,m=1$ Skyrmion on gravity has been studied before
\cite{Luckock:1986tr,Droz:1991cx,Bizon:1992gb}, this branch of gravitating Skyrmions is linked to the flat space solutions.
Similarly, when gravity is coupled to the Skyrme model, a branch of gravitating axially-symmetric
Skyrmion–-anti-Skyrmion chains, which are formed from $m$ constituents of
charge $±2$, emerges smoothly from the corresponding flat space sphaleron solutions.

However, $|n| = 1$ sphalerons of the massless Skyrme model do not exist in the flat space \cite{Krusch:2004uf}.
Clearly, coupling to gravity yields additional attractive interaction between the constituents.
We have found numerical evidence that in the Einstein-Skyrme model
a branch of self-gravitating charge 1 Skyrmion and
charge -1 anti-Skyrmion pairs emerges at the critical non-zero value of the gravitational
coupling $\alpha_0=0.08$ and extends
up to a maximal value $\alpha_{cr}=0.251$ where it merges the upper mass branch.
For the $|n| = 1$ S-A-S chain the critical values of the gravitational coupling are $\alpha_0=0.082$ and
$\alpha_{cr}=0.2305$, respectively. Along this so-called Skyrmion branch,
the distance between the constituents, which is defined as
the points in space where the component of the Skyrme field $\phi_3$ is equal to its anti-vacuum value $-1$,
is slowly decreasing.

As the gravitational coupling constant increases, the background is getting more and more deformed
and, at some critical value of the coupling, when the  gravity becomes too strong for solutions to
persist, the Skyrmion branch merge with branch of different type, cf Figs.~\ref{f-1}-\ref{f-3}.
The critical value $\alpha_{cr}$, at which a backbending is observed,
slightly decreases as the number of the constituents of the Skyrmion–-anti-Skyrmion chains is increasing.

The parametrization \re{metric} allows us to find the dimensionless ADM mass of the configuration $M$, it is defined by
the value of the derivative of the metric function $f$ at the boundary
\be
\label{mass-1}
M = \frac{1}{2 \alpha^2}\lim_{x\to\infty} \partial_x f
\ee
In order to perform another check of our numerics for correctness, we compare this value with the results of
direct evaluation of the integral over the $T_{00}$ component of the total energy-momentum tensor \re{energy}.

Along the first (lower) branch the mass of the gravitating
Skyrmion--anti-Skyrmion chains decreases with increasing $\alpha$, since with increasing
of gravitational strength the attraction in the system increases.
Along the second (upper) branch, in contrast, the mass \re{mass-1}
increases strongly with decreasing the coupling $\alpha$, and the solutions
shrink correspondingly. However, in the limit of vanishing coupling
constant when the mass $M$  diverges, the pattern of evolution of the configurations depends on the
value of integer numbers $n$ and $m$.

As shown in \cite{Bizon:1992gb} the spherically symmetric $n=1$, $m=1$ configuration in this
limit approaches the lowest mass spherically symmetric Bartnik-McKinnon solution \cite{Bartnik:1988am}.
The structure of the limiting $F_\pi \to 0$ configuration can be better understood
when we introduce the rescaled radial coordinate $\hat x = x/\alpha$ and the rescaled mass $\hat M = \alpha M $
\cite{Bizon:1992gb}. Then the reparametrization of the Skyrmion field
\be
\phi^1(\hat x) = \sqrt{1-\omega(\hat x)^2}\sin\theta; \qquad
\phi^2(\hat x) = \sqrt{1-\omega(\hat x)^2}\cos\theta; \qquad
\phi^3(\hat x)=\omega(\hat x)
\ee
allows us to represent the system of the field equations on the function $\omega(\hat x)$ and the metric functions $f(\hat x),
l(\hat x)=m(\hat x)$ in the form, which is identical to the system of the $SU(2)$ Einstein-Yang-Mills equations for the
Bartnik-McKinnon configuration with the gauge connection $A_i^a=\varepsilon_{ian}(1-\omega(\hat x)) {\hat x}_n/2$
\cite{Bartnik:1988am}. On the other hand, the rescaled mass of the axially symmetric charge $n=2$ Skyrmion is higher,
the limiting configuration on the corresponding upper branch is the generalized BM solution
\cite{Ibadov:2004rt,Kleihaus:1996vi}.

Let us now consider axially-symmetric $m=2$ Skyrmion–-anti-Skyrmion pairs. This configuration resides in the vacuum sector,
these chains exhibit an analogous dependence on the coupling constant $\alpha$ as the $m=1$ Skyrmions.
We found that this self-gravitating $n=1$ and $n=2$ S-A sphaleron with decreasing $\alpha$ on the upper branch
also smoothly approaches the corresponding generalized BM solution,
as illustrated in Figs.~\ref{f-1},\ref{f-2}.

To clarify this observation, we plot in Fig.\ref{f-4} the metric function $f$ and
the third component of the Skyrme field for the $n=1$ configurations
for a small value of the gravitational coupling $\alpha$ on the
upper branches.

Clearly, we can identify three distinct regions. As seen in Fig.~\ref{f-4}, right column,
in the first region the metric function $f$ remains very small but constant without any significant
angular dependency. In the second transition region, the metric varies up to upper value $f=1$,
this is a small region where the energy of the matter field of the S-A configuration is located.
Finally, in the third outer region, the metric functions are approaching the flat space
limiting values.

This pattern is similar to the known picture of
the evolution of self-gravitating  monopole-antiminopole pairs \cite{Kleihaus:2000hx,Kleihaus:2004fh},
which on the upper unstable branch are also linked
to the generalized Bartnik-McKinnon limiting solutions. One might expect the same analogy holds for the
Skyrmion--anti-Skyrmion gravitating chains and monopole-antimonopole chains
with higher number of constituents. However, we observe a bit different pattern for the coupling constant dependence
of the S-A-S system on the upper branch.
For some small critical value of the gravitational constant $\alpha$ the
central component of the chain, the anti-Skyrmion of charge $-n$, is located into the interior
region while remaining in the transition region pair of Skyrmions with positive charges $n$
becomes separated since the weak gravitational interaction there
cannot stabilize it. Thus, the $n=1,2$ S-A-S chains are broken at $\alpha_{cr}^{(up)} = 0.06$ and
$\alpha_{cr}^{(up)} = 0.075$, respectively.

The composition of the $n=2, m=3$ solution at $\alpha^2 = 0.01$ is exhibited in
Fig.~\ref{f-5}, in Fig.~\ref{f-4}, bottom row,  we presented the the metric function $f$ and
the third component of the Skyrme field for the $n=1$ S-A-S configuration.

However,
one may expect that for higher values of the components of the chains, say for $m=4$
S-A-S-A configuration with $|n| \ge 3$,
will stay unbroken along the upper branch all the way down to the limit $\alpha \to 0$.
This conjecture is based on the preliminary results
we found considering $m=4$, $n=3$ solution, they suggest that as $\alpha$ decreases,
the location of the inner S-A pair
moves continuously inwards along this branch and tends to the origin in the limit
$\alpha \to 0$.
However, the attractive interaction between the remaining
in the outer flat space region S-A pair could be strong enough to overcome the repulsive force.
The location of the outer S-A pair in this case
approaches a finite value, which is in a good agreement with the separation
of the corresponding S-A pair in the flat space. Thus, similar to the case of $m=4$ monopole-antimonopole chain,
the resulting configuration may be thought of as composed of a scaled (generalized) BM solution in
the inner region and a flat space soliton--anti-soliton pair solution in the outer region.

\begin{figure}[hbt]
\lbfig{f-5}
\begin{center}
~~~~
\includegraphics[height=.32\textheight, angle =-90]{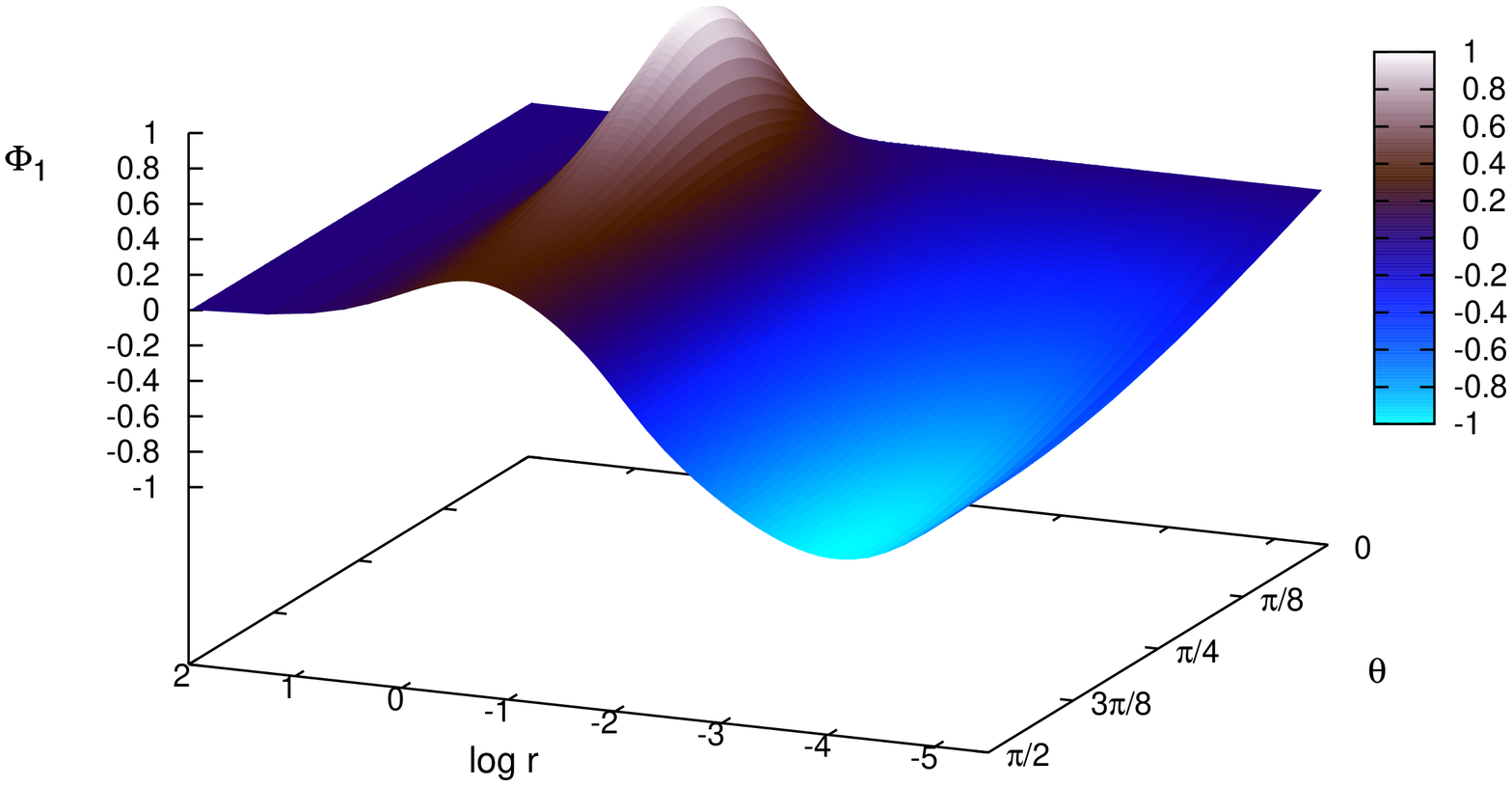}
\includegraphics[height=.32\textheight, angle =-90]{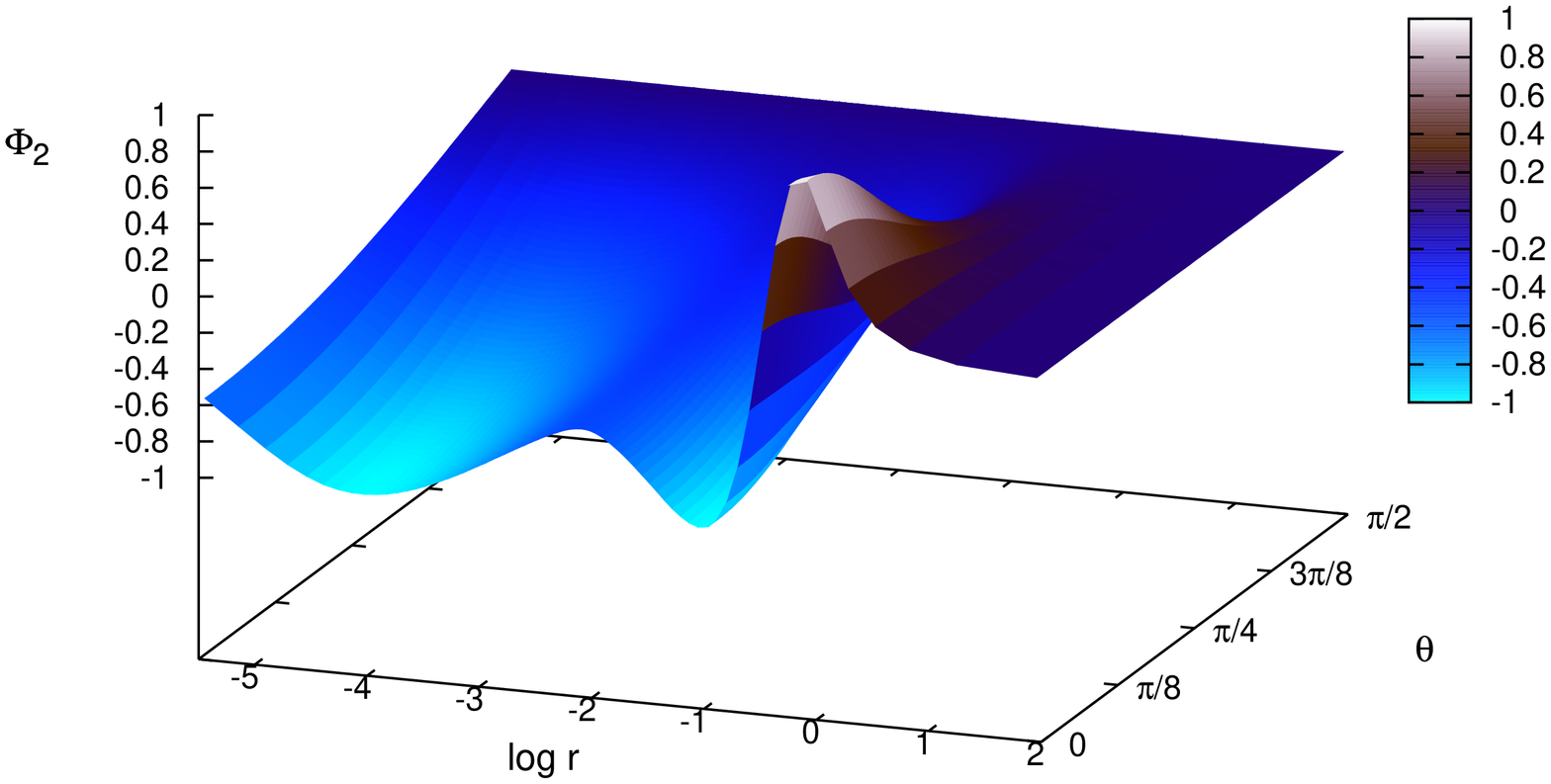}
\includegraphics[height=.32\textheight, angle =-90]{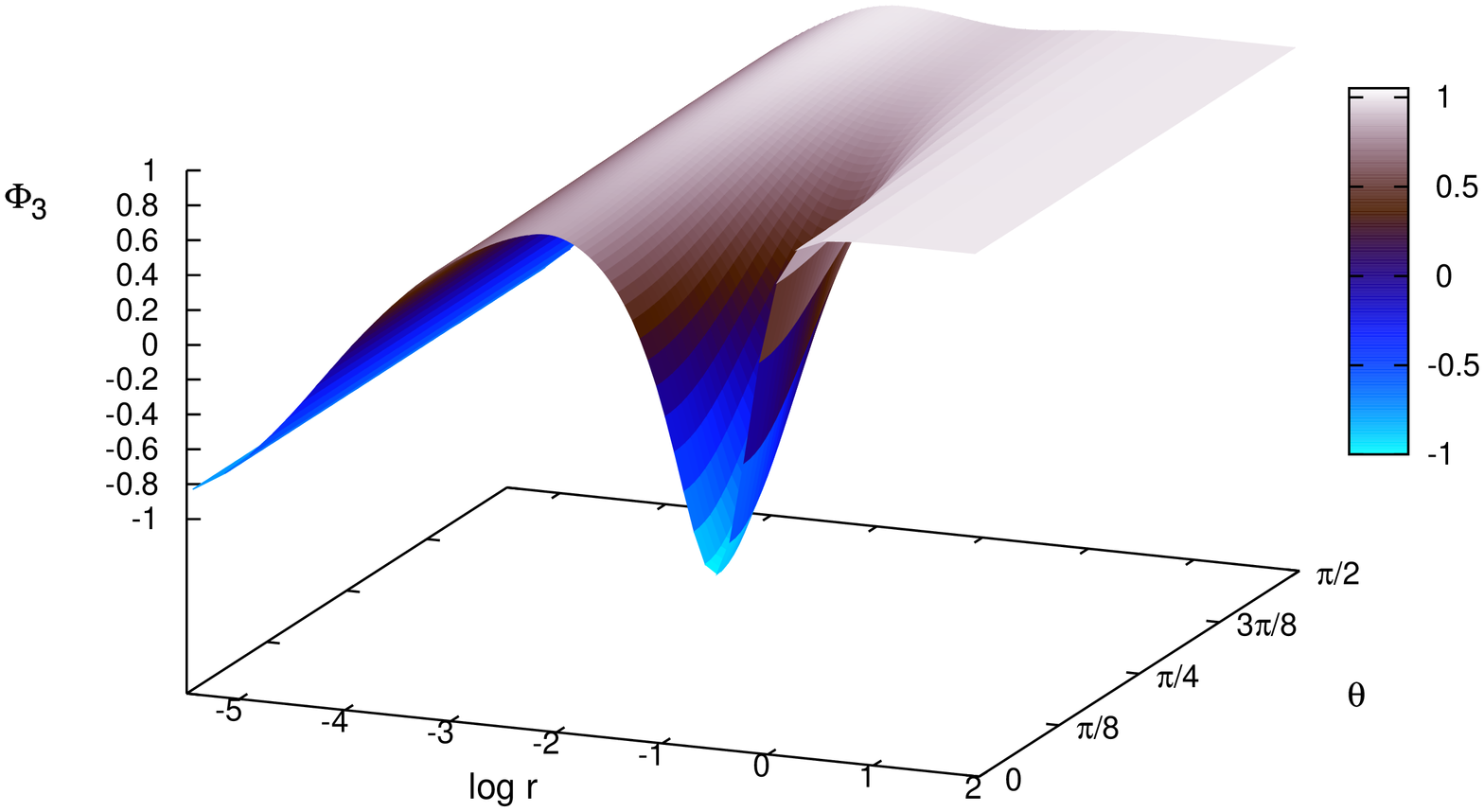}
\end{center}
\caption{\small
The components of the field of the
gravitating $n=2,m=3$ S-A-S chain  on the upper branch are shown at $\alpha^2 = 0.01$
in the logarithmic radial coordinate $\log r$.
}
\end{figure}

\section{Conclusions}
The main purpose of this work was to present new type of self-gravitating sphaleron solitons in the Einstein-Skyrme theory.
The configurations of that type are characterized by two integers $n$ and $m$, where $|n|$ is the
winding numbers of the constituents, and the
second integer $m$ defines the type of the configuration, it is a deformation of the topologically trivial secror
for even $m$ and for odd values of $m$ the Skyrmion–anti-Skyrmion chain is a deformation of the axially-symmetric Skyrmion of
degree $n$ \cite{Shnir:2009ct}.

In the present work we have focussed on the Skyrmion--anti-Skyrmion chains with
2 and 3 constituents, S-A and S-A-S, respectively.
Winding
number of each individual component is restricted to lower values $n=1,2$. These solutions
are asymptotically flat, and globally regular.

Concerning the dependence of the gravitating Skyrmion--anti-Skyrmion chains on the gravity coupling constant,
we generally observe the pattern which is similar yet slightly different from the picture observed previously
in the case of the sphaleron solutions of the Einstein-Yang-Mills-Higgs theory. First, in the absence of the
pion-mass potential,
there are branches of self-gravitating $|n| = 1$ Skyrmion-–anti-Skyrmion chains which
emerge at some critical non-zero value of the gravitational coupling, these branches do not have flat space limit.
In contrast, the branches of self-gravitating $|n| \ge 2$ Skyrmion-–anti-Skyrmion chains
emerge from the corresponding flat space configurations.

In both case these lower branches merge at some maximal
value of the effective gravitational coupling the upper branches.
The pattern of evolution along these branches is related with effective decreasing of the Skyrme coupling constant
$F_\pi$ and the structure of the limiting configurations depends on the type of the configuration.
The branches of gravitating $m=1$ Skyrmions and $m=2$ S-A pairs extends all the way back to the limit
$\alpha\to 0$ where solutions approach the corresponding (generalised) Bartnik-McKinnon solutions of the
$SU(2)$ Einstein-Yang-Mills theory. However, the upper branch of gravitating S-A-S chain exist up to some critical
value of the gravitational coupling at which the chain becomes broken. Then the  anti-Skyrmion
of degree $n$ remains constrained at the interior region whereas the pair of identically charged Skyrmions
in the outer region becomes separated.

Here it is interesting to compare the results with the pattern of evolution of the self-gravitating
solitons of the Einstein-Faddeev-Skyrme model \cite{Shnir2015}.
The structure of the Skyrme model and the Faddeev-Skyrme model look similar,
the corresponding Lagrangians, like \re{Skyrme}, include the usual sigma model term, the Skyrme term, which is quartic in
derivatives of the scalar field $\phi^a$, and the optional potential term which does not contain the derivatives.
However the topological properties of the corresponding solitons are different,
the finite energy solutions of the Faddeev-Skyrme
model, the Hopfions, correspond to the map $\phi^a:\mathbb{R}^3 \to S^2$ which belongs
to an equivalence class characterized by the third homotopy group $\pi_3(S^2)=\mathbb{Z}$ \cite{Faddeev}.
Thus, the corresponding integer topological invariant associated with the triplet of scalar fields $\phi^a$
constrained to the unit sphere $S^2$ is
known as the Hopf number $Q$. This invariant can be interpreted geometrically as
the linking number of two loops obtained as the preimages of any two generic distinct points
on the target space $S^2$.

Similar to the case of the Einstein-Skyrme model considered above, we can couple the scalar triplet of the fields of the
Faddeev-Skyrme model to gravity. Considering the gravitating static axially symmetric Hopfions of lower degree $Q\le 4$
we observe the same general pattern as for $m=1$ Skyrmion solutions  of the Einstein-Skyrme model, S-A chains
and the monopole-antimonopole chains of the
Einstein-Yang-Mills-Higgs theory.
We have found numerical evidence that, when gravity is coupled to the Fadeev-Skyrme model,
a branch of gravitating Hopfions emerges from the flat space Hopfion solution and extends
up to a maximal value $\alpha_{cr}$ where it merges the upper mass branch, cf
Fig.~\ref{f-6} vs Figs.~\ref{f-1},\ref{f-2}.
\begin{figure}[hbt]
\lbfig{f-6}
\begin{center}
\includegraphics[height=.32\textheight, angle =-90]{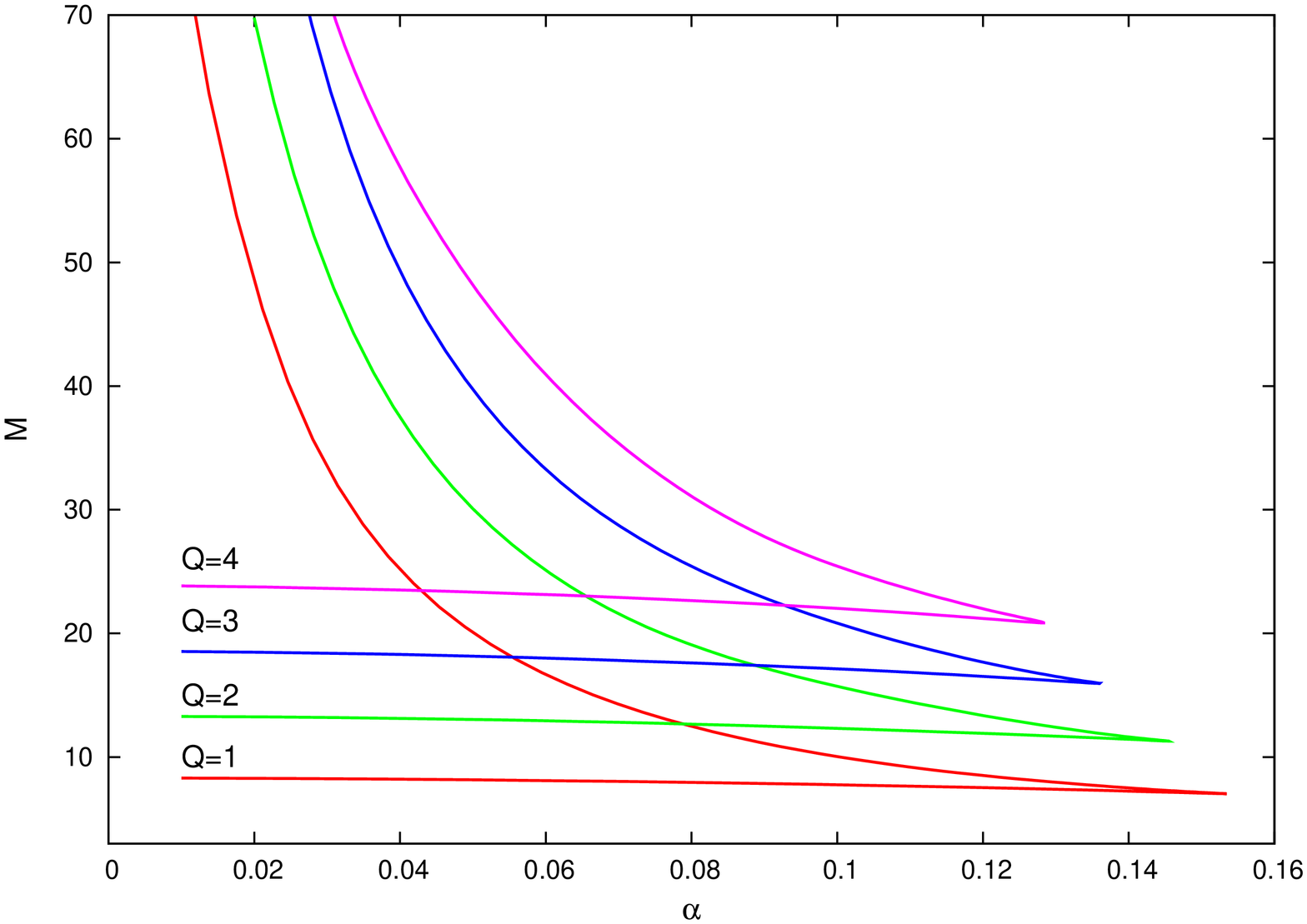}
\includegraphics[height=.32\textheight, angle =-90]{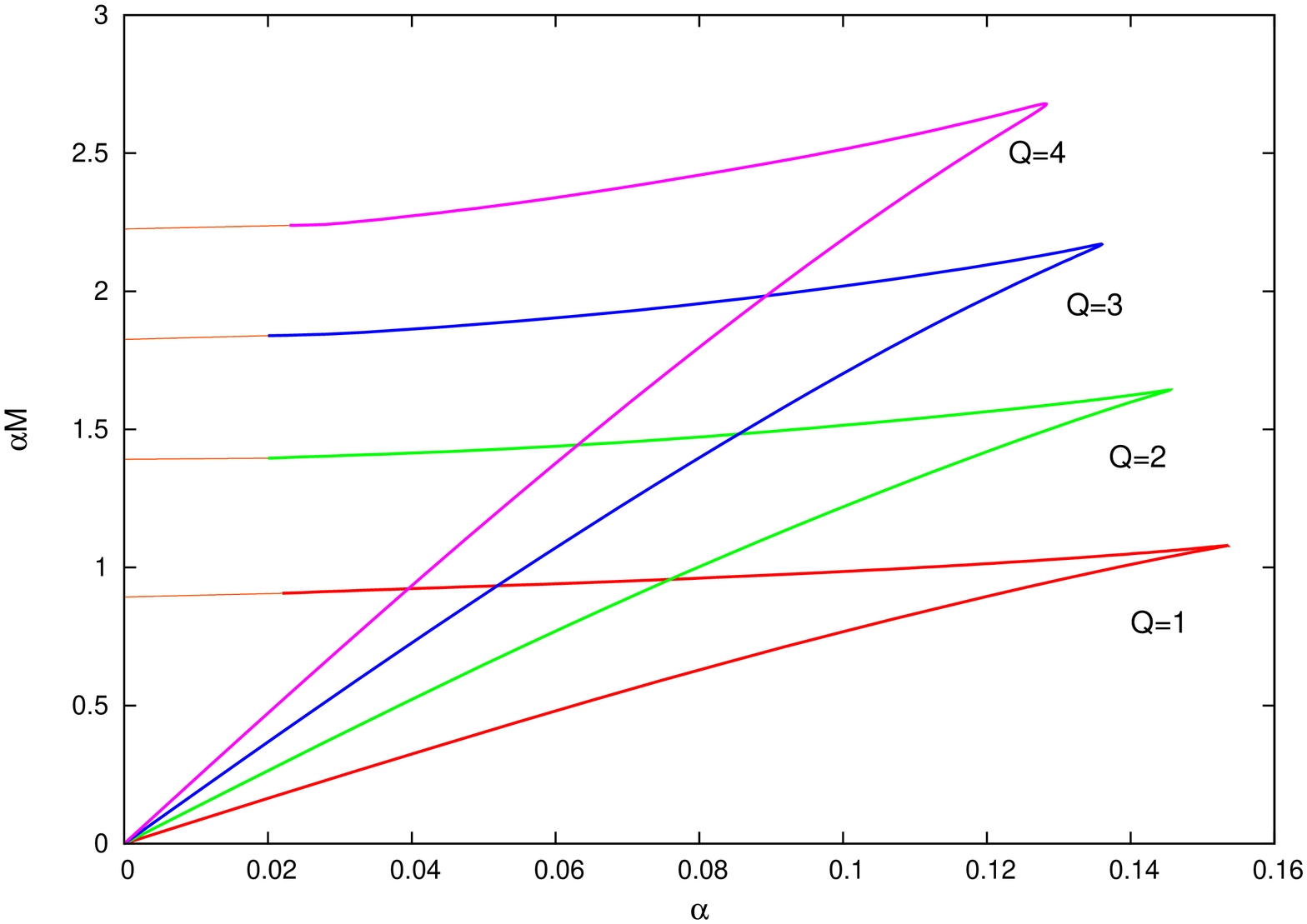}
\end{center}
\caption{\small
The mass $\mu$, and the scaled mass $\hat \mu$ of the gravitating  Hopfions of degrees $1$ to $4$
are shown as functions of the coupling constant $\alpha$. The thin lines extend the Hopfion
curves of the scaled mass to the mass of the corresponding generalized Bartnik-McKinnon solution.
}
\end{figure}
Once again, the upper branch extends back to the limit $\alpha \to 0$
where solutions approach the corresponding (generalised) Bartnik-McKinnon solutions of the
$SU(2)$ Einstein-Yang-Mills theory, as seen in Fig..~\ref{f-6}, right plot.

We hope to return elsewhere with a discussion of some of these interesting aspects of the self-gravitating Hopfions.

There are various possible extensions of the solutions discussed in this work. Clearly, it would be interesting to investigate
how inclusion of the pion mass term may affect the results. Also, we expect the $m\ge 4$ Skirmion--anti-Skyrmion chain
configurations with constituents of higher degree $|n|\ge 3$ may stay unbroken on the upper branch as $\alpha \to 0$.
Finally, our preliminary
results indicate the existence of static axially symmetric black hole solutions
with Skyrmion's hair, consisting of chain-like structures.
It would be also interesting to address the
question how inclusions of a cosmological constant will affect the
properties of a gravitating sphalerons of the Skyrme model.

\section*{Acknowledgements}
I thank Olga Kichakova, Jutta Kunz, Eugen Radu and Michael Volkov for
useful discussions and valuable comments. I am gratefully acknowledge
support from the A.~von Humboldt Foundation in
the framework of the Institutes Linkage Programm.

\end{document}